\documentclass[aps,twocolumn,amsmath,amssymb,superscriptaddress]{revtex4}

\usepackage[dvipdfmx]{graphicx}
\usepackage{dcolumn}
\usepackage{bm}
\usepackage{color}
\usepackage{amstext,amssymb}
\newcommand{\be}{\begin{equation}}
\newcommand{\ee}{\end{equation}}
\newcommand{\bea}{\begin{eqnarray}}
\newcommand{\eea}{\end{eqnarray}}
\newcommand{\lsim}{\raise.35ex\hbox{$<$}\kern-0.75em\lower.5ex\hbox{$\sim$}}
\newcommand{\gsim}{\raise.35ex\hbox{$>$}\kern-0.75em\lower.5ex\hbox{$\sim$}}
\usepackage[normalem]{ulem} 

\usepackage[
colorlinks=true,
citecolor=blue,
urlcolor=blue,
setpagesize=false]{hyperref}
\begin{document}
%
%
\title{Quantum Monte Carlo study of superfluid density in quasi-one-dimensional hard-core bosons: 
Effect of suppression of phase slippage}
\author{A. Masaki-Kato}
\affiliation{Computational Condensed Matter Physics Laboratory, RIKEN Cluster for Pioneering Research (CPR), Wako, Saitama 351-0198, Japan}
\affiliation{Computational Materials Science Research Team, RIKEN Center for Computational Science (R-CCS),  Kobe, Hyogo 650-0047,  Japan}
\author{S. Yunoki}
\affiliation{Computational Condensed Matter Physics Laboratory, RIKEN Cluster for Pioneering Research (CPR), Wako, Saitama 351-0198, Japan}
\affiliation{Computational Materials Science Research Team, RIKEN Center for Computational Science (R-CCS),  Kobe, Hyogo 650-0047,  Japan}
\affiliation{Computational Quantum Matter Research Team, RIKEN, Center for Emergent Matter Science (CEMS), Wako, Saitama 351-0198, Japan}
\author{D. S. Hirashima}
\affiliation{Department of Natural Sciences, International Christian University, Mitaka, Tokyo 180-8585, Japan}%

\date{\today}
%
\begin{abstract}
 We study the superfluid density of hard-core bosons on quasi-one-dimensional lattices 
 using the quantum Monte Carlo method.
Because of phase slippage, 
the superfluid density drops quickly to zero at finite temperatures with increasing the 
system length $\ell$ and the superfluid transition temperature is zero in one 
spatial dimension 
and also in quasi-one dimension in the limit of $\ell\rightarrow\infty$. 
We calculate the superfluid density of a model where no phase 
slippage is allowed and show 
that the superfluid density remains finite at finite temperatures
even in the one-dimensional limit. 
We also discuss how finite superfluid density can be observed in a quasi-one-dimensional system 
using a torsional oscillator.
\end{abstract}







\maketitle

\section{Introduction\label{sec:intro}}
Superfluidity is one of the most fascinating phenomena
in condensed matter systems. It is well know that the spatial dimensionality
plays a remarkable role in superfluidity. In two dimensions, the 
Berezinskii-Kosterlitz-Thouless (BKT)
transition~\cite{Berez,KT1}, a unique
topological phase transition, occurs at a finite
temperature despite the absence of
true long-range order~\cite{Bert}. 

Recently, $^{4}$He atoms confined in straight nanopores have attracted much 
attention~\cite{Wada0,Tani0,Ike,Toda,Tani1,Tani2,Tani3,Wada3,Wada4,Tani4,Tani5,TaniGlyde,Wada5,Gerv}.
For example, 
Wada and coworkers studied He atoms adsorbed on the inner walls of
one-dimensional pores of porous material, 
FSM (Folded sheet mesoporous materials)-16~\cite{Wada0,Tani0,Ike,Toda};
typical pore length is 200-300 nm and its diameter $R$ can be
systematically changed from 
$R=1.5$ nm to $4.7$ nm. 
Superfluid density was measured using a torsional oscillator
and a frequency shift was found to set in at a temperature
close to the 
BKT transition temperature, $T_{\rm BKT}$~\cite{Berez,KT1}, 
determined by the areal density of adsorbed He atoms.
They carefully analyzed the results and concluded that
the frequency shift was caused by finite superfluid density
in the one-dimensional part of the system, i.e., 
the one-dimensional He tube~\cite{Ike,Toda}.

On the other hand, Taniguchi and Suzuki studied
superfluidity of liquid $^{4}$He filling
nanopores~\cite{Tani1,Tani2,Tani3,Tani4,Tani5}. 
Superfluid density was then measured with a torsional
oscillator. They found a two-step increase, 
that is, they found an additional increase in the
resonance frequency
at a temperature lower than the bulk $\lambda$ transition
temperature. They also ascribed the second increase to the onset of
superfluidity of liquid He filling one-dimensional 
pores~\cite{Tani1,Tani2,Tani3}.

In these experiments, three-dimensionality
plays only a minor role in contrast to the previous
experiments using interconnected porous 
materials~\cite{Bert,Crook,Adams,Beamish,Lie,Yamamoto},
because the pores are connected only via their ends. Moreover,
as the pores are regularly arranged, randomness caused by 
irregular connection of pores, which has significant effects
in Vycor glass, for example, is also
considered to be irrelevant.

It is a
well-established fact that no Bose-Einstein condensation
(BEC) occurs at a nite temperature in one or two
dimensions~\cite{ Hohenberg, Mermin-Wagner}.
However, the relation of existence or absence of BEC
to superuidity is not necessarily a well-understood problem\cite{Lieb}.
In particular, we should note that superfluidity is detected 
dynamically in torsional oscillator experiments.
Therefore, in discussing superfluidity in one
dimension or quasi-one dimension, a dynamical aspect of the phenomenon
has to be considered.
For example, Shevchenko
showed that the characteristic temperature for
one-dimensional superfluidity is given by
\begin{equation}
T_{\rm c} \sim \frac{\hbar^{2}n_{1}}{k_{\rm B}M\ell_{z}},
\end{equation}
where $n_{1}$ is the one-dimensional number density of boson
atoms, $M$ the atomic mass, and $\ell_{z}$ the one-dimensional
length of the system~\cite{Shev}. Although it 
simply vanishes as 
$\ell_{z}\rightarrow\infty$,
he argued that 
superfluidity would be observed at a much higher 
temperature than $T_{\rm c}$,
if $\omega \tau \gg 1$ is fulfilled, where $\omega$ is the
frequency at which superfluidity is measured
(e.g., the frequency of a torsional oscillator)
and $\tau$ is the relaxation time of supercurrent~\cite{Shev}.

A more explicit argument was independently
given by Machta and Guyer~\cite{MG,Machta}.
They proposed two different definitions of
superfluid density. One is denoted by 
$\rho_{\rm s}$, 
which
is the coefficient of the increase in the free energy
in the presence of slow supercurrent,
and the other by $\rho_{\rm p}$, which
is the coefficient of the increase in the free energy
caused by an  infinitesimal phase twist between both
ends of the system.
They found the relation
\begin{equation}
{\rho}_{\rm p}(T) \simeq 
2 L_{\rm eff} \frac{k_{\rm B}T}{{J}}
\exp{\left[-\frac{L_{\rm eff}}{2{\rho}_{\rm s}(T)}
\frac{k_{\rm B}T}{{J}}\right]},
\label{eq:MG}
\end{equation}
at temperature $T \gg
J{\rho}_{\rm s} /(k_{\rm B}L_{\rm eff})$, where
$L_{\rm eff}=\ell_{z}/(n_{1}a^{2})$, ${J}=
\hbar^{2}/(Ma^{2})$, and $a$ is the average
interparticle distance~\cite{Machta}. 
The superfluid densities are
normalized so that ${\rho}_{\rm s,p}(T=0)=1$. 
Out of the two superfluid densities,
${\rho_{\rm p}}(T)$ is affected by phase slippage and 
readily vanishes at
$T \gtrsim
{J}{\rho}_{\rm s}/(k_{\rm B}L_{\rm eff})
\simeq
\hbar^{2}n_{1}/(k_{\rm B}M\ell_{z})\sim T_{\rm c}$. 
In other words, it vanishes at any finite temperatures
in the limit of $\ell_{z}\rightarrow\infty$.
On the other hand,
${\rho}_{\rm s}(T)$ does not
suffer from phase
slippage and can be finite at finite temperatures.
They argued that it is ${\rho}_{\rm s}(T)$ 
that is 
observed in torsional oscillator experiments,
but did not discuss the explicit temperature dependence of
$\rho_{\rm s}(T)$.
A similar relation between $\rho_{\rm s}(T)$ and
$\rho_{\rm p}(T)$  
was also derived by Prokof'ev and Svistunov~\cite{SP}. 
It should also be noted that it is $\rho_{\rm p}(T)$ that is
obtained with calculations under the thermal equilibrium
condition.

It is not trivial which one, $\rho_{\rm s}(T)$
or $\rho_{\rm p}(T)$, will be observed in a torsional
oscillator experiment. If $\omega\tau \gg 1$ at low temperatures, it must be
$\rho_{\rm s}(T)$ that will be observed in an experiment
as was suggested by Machta and Guyer~\cite{Machta}. 
However, $\tau$ is temperature dependent and should
decrease as $T$ increases. Therefore, 
$\rho_{\rm p}(T)$ will be observed once $\omega\tau$ becomes
small enough. In actual dynamical experiments, this kind
of crossover will be observed. 

Superfluid density in one-dimensional systems was also analyzed
using the Tomonaga-Luttinger theory~\cite{Bon,Maestro,Caza}.
In particular, the dynamical aspect of superfluid 
behavior
was discussed by Eggel {\it et al}.~\cite{Caza}
and experimental results~\cite{Tani5} 
were analyzed based on this theory.
However, the theory is limited to low temperatures and 
the temperature range or the range of pore radius 
where the theory can be justified is not clear.

Quantum Monte Carlo simulations were also performed for
liquid $^{4}$He confined in nanopores \cite{Vr1,Vr2}.
The results, in particular, those in narrower pores,
$R < 0.4$ nm with $R$ being the radius of a pore, 
are successfully analyzed using the
Tomonaga-Luttinger theory. However, for wider pores,
$R > 0.9$ nm, 
the system length used in the simulations
may not be long enough to study the quasi-one-dimensional
cases.

Superfluid density in quasi-one-dimensional
systems was also analyzed on the basis of classical spin
models ($XY$ models)~\cite{Yama1,Yama2}. 
Superfluid density $\rho_{\rm s}(T)$
that is not affected by phase slippage in quasi-one dimension
was calculated using a special boundary condition or
a restricted sampling method~\cite{Yama2}. It was then found that,
without the effect of phase slippage, superfluid density
can survive up to the transition temperature of the extended
film or the bulk system even in the one-dimensional
limit~\cite{Yama2}. 
Although the main conclusion in Ref.~\onlinecite{Yama2} is expected to 
be also valid in quantum systems, it is highly desirable to demonstrate 
it explicitly in a
quantum system. This is precisely the purpose of this paper.

In this study, we examine superfluid density of hard-core bosons on 
quasi-one-dimensional lattices using the quantum Monte Carlo method. 
As was done in Ref.~\onlinecite{Yama2}, we calculate superfluid density
$\rho_{\rm s}(T)$ by modifying the model used in the calculation. 
In this study, we suppress phase slippage by introducing special transfer 
integrals. We then show that superfluid density can remain finite
up to the BKT transition temperature $T_{\rm BKT}$ or the bulk
transition temperature $T_{\lambda}$ even in the one-dimensional limit
when the effect of phase slippage is completely suppressed.

The rest of this paper is organized as follows.
Section~\ref{sec:model} introduces 
a hard-core Bose-Hubbard model and modify it so that
the phase slippage is prevented.
In addition, we define the superfluid density for this modified
model.
Section~\ref{sec:results} presents the results of the simulations, 
which
clearly show that the superfluid density can be finite at 
high temperatures when the phase slippage is not allowed.
Section~\ref{sec:conclusion} summarizes this paper.

\section{Model and method}\label{sec:model}

\subsection{Model without phase slippage}

In order to study a quasi-one-dimensional system such as $^4$He atoms in nanopores,
we consider hard-core bosons on an anisotropic square 
or cubic lattice described 
by the following Hamiltonian: 
\begin{equation}
{\cal H}=-\sum_{\langle i,j\rangle}\left( t_{ij}b^\dagger_ib_j + {\rm H.c.}\right),
\label{eq:HBHM}
\end{equation}
where $b_i$ $(b_i^\dagger)$ is the annihilation (creation) operator of a boson
at site $i$ and no multiple occupancy at the same site is allowed because of the strong
repulsion between bosons, i.e., 
$b_{i}^{\dagger}b_{i}=0$ or $1$.
For simplicity, we consider only the transfer integral $t_{ij}$
between the nearest neighboring sites, and accordingly 
the sum in Eq.~(\ref{eq:HBHM}) runs over all 
the nearest neighboring sites $\langle i,j \rangle$. 
Furthermore, we do not consider 
the interaction between bosons at different sites. 
In this study, we set the boson density at half filling, i.e.,  
$N=0.5 N_{\rm L}$, where $N$ ($N_{\rm L}$) is the total number of bosons (lattice sites), 
and thus the chemical potential $\mu$
is always zero. 
Note that this model can be mapped onto the spin $S=1/2$ $XY$ model with only
the nearest-neighbor exchange interaction with no
external magnetic field~\cite{MatsuMatsu}.

\begin{figure}[htpb]
  \includegraphics[angle=0,width=8.5cm,trim= 0 0 0 0,clip]{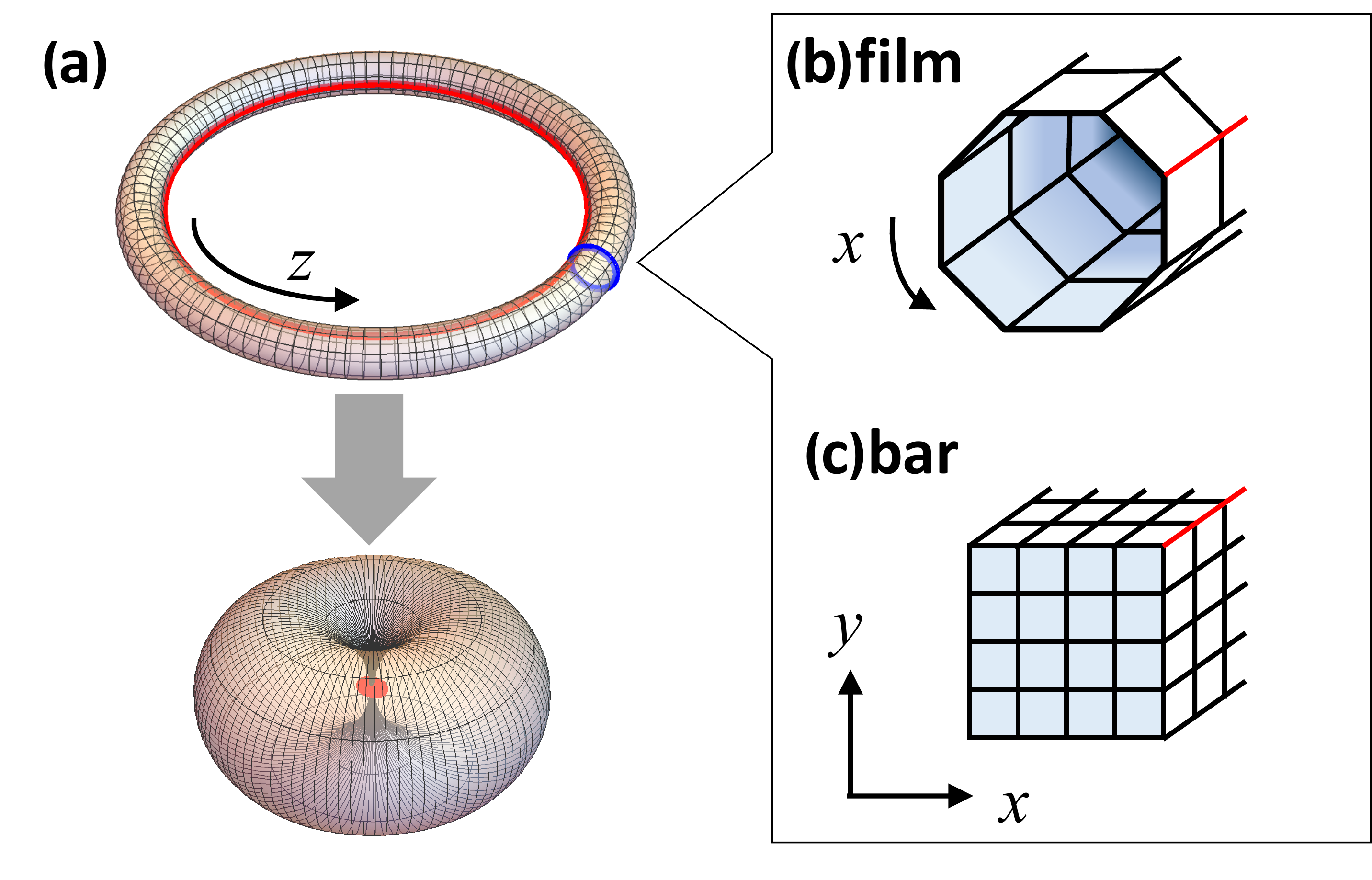}
\caption{
(a) Quasi-one-dimensional lattice system with the periodic boundary condition 
along 
the $z$ direction, forming a torus-like geometry.
Phase slippage is suppressed in a spin model if the central hole of the torus is closed~\cite{Yama2}. 
This is equivalent to setting the transfer integrals along a single row 
(indicated by the red line) 
to be infinite in the hard-core boson model studied here.
The inner part is empty in the film case (b) and it is filled with lattice sites in the bar case (c). 
}
\label{fig:systems}
\end{figure}

To simulate $^{4}$He atoms adsorbed on the 
inner walls of nanopores~\cite{Wada0,Ike,Toda,Wada3,Wada4,Wada5}, we consider 
an anisotropic two-dimensional square lattice, i.e., a film, 
composed of $L_{x}\times L_{z}$ sites with $L_{z} \gg L_{x}$ (see Fig.~\ref{fig:systems}).
The periodic condition is imposed in both directions. 
We thus consider hard-core bosons on a long tube, as schematically shown in Fig.~\ref{fig:systems}(b).
For a film, the effective length $L_{\rm eff}=\ell_{z}/(n_{1}a^{2})
\sim L_{z}/L_{x}\sim \ell_{z}/\ell_{x}$, i.e., the aspect ratio of the anisotropic lattice, 
because $n_{1}=N/\ell_{z}=N_{\rm L}/(2\ell_z)$ and $L_{\alpha}\sim \ell_{\alpha}/a$
($\alpha=x$ and $z$).
Experimentally, the aspect ratio $L_{z}/L_{x}$ can be estimated to be 
$15-50$~\cite{Wada0,Ike,Toda,Wada3,Wada4,Wada5}. On the other hand, to simulate $^{4}$He atoms filling nanopores,
we consider an anisotropic three-dimensional cubic lattice, i.e., a bar, 
composed of $N_{\rm L}=L_{x}L_{y}L_{z}$ sites with $L_{z} \gg L_{x}, L_{y}$ 
[see Figs.~\ref{fig:systems}(a) and \ref{fig:systems}(c)].
For a bar, the effective length $L_{\rm eff}$ can be estimated as $L_{\rm eff}
\sim L_{z}/(L_{x}L_{y})
\sim \ell_{z}a/(\ell_{x}\ell_{y})$ and 
typically 
$L_{z}/(L_{x}L_{y}) = 5-30$~\cite{Tani0,Tani1,Tani2,Tani3,TaniGlyde}.
In our simulations, the periodic boundary condition is imposed in 
the $z$-direction and the open boundary 
condition is applied in the remaining two directions.

The superfluid density $\rho_{\rm p}(T)$
in these systems vanishes at finite temperatures in the
limit of $L_{z}\rightarrow \infty$ because of phase slippage. 
To calculate the superfluid density $\rho_{\rm s}(T)$ that 
is not affected by phase slippage, a slight modification of 
the model is required. 
In a spin model, phase slippage is suppressed when all the $L_{z}$ spins in a single row 
of the lattice (out of $L_{x}$ or $L_{x}\times L_{y}$ rows) are replaced with a single spin~\cite{Yama2}. 
By doing so, one can close
the central hole of the torus as is shown in Fig.~\ref{fig:systems}(a). 
This is equivalent to setting the exchange interaction
in this row to be infinity. The exchange interaction in the 
$XY$ model is mapped to the transfer integral in the hard-core boson system studied here.
Thus, by setting the transfer integral in a single
row of the lattice (out of $L_{x}$ or $L_{x}\times L_{y}$ rows) along the $z$ direction to be infinity, we can 
prohibit phase slippage.
With this modification,
the translational invariance along the
transverse directions ($x$- or/and $y$-direction) is violated, but 
the system remains translational invariant in the $z$ direction.

In the numerical calculations, 
we set the transfer integral $t_{ij}$ along this single row to be $t^{*}$, which is much larger 
than the other transfer integral $t_{ij}=t$. Typically, we set
$t^{*} \sim L_{z} t$ but we also investigate the dependence of the
results on the choice of $t^{*}$.
As is shown in the following, by calculating the superfluid density of the 
model with this transfer integral $t^{*}$,
we can obtain the superfluid density $\rho_{\rm s}(T)$ that is not affected
by phase slippage.

\subsection{Method: superfluid density}

To calculate superfluid density,
we apply the worldline Monte Carlo method employing the directed-loop 
implementation~\cite{DLA,Kawashima,Kato} of the worm algorithm~\cite{PST}.
The well-known definition of the superfluid density~\cite{Pollock} in a 
spatially 
homogeneous system is given by
\begin{equation}
\rho_s=\frac{
\sum_\alpha \langle L^2_\alpha  W^2_\alpha \rangle
}{2t\beta N_{\rm L}},
\label{eq:sfd}
\end{equation}
where ${\boldsymbol L}=(L_{x}, L_{y}, L_{z})$ stands for
the linear system size in a three-dimensional cubic lattice system, 
${\boldsymbol W}=(W_{x}, W_{y}, W_{z})$ is the 
winding number in each spatial direction, 
$\langle\cdots\rangle={\rm Tr}\left( e^{-\beta{\cal H}}\cdots \right)/{\rm Tr}e^{-\beta{\cal H}}$,
$\beta=1/T$, and the Boltzmann constant $k_{\rm B}$ is set to be 1.
The total number of lattice sites is $N_{\rm L}=L_xL_yL_z$. 
The hopping integral $t$ in Eq.~(\ref{eq:sfd}) is assumed to be
uniform. 
When we calculate the winding number $W_\alpha$ $(\alpha=x,y,$ and $z$), 
we count the number of kinks of worldlines that correspond to the hopping operator $b^\dagger_ib_j$. 
For example, the winding number in the $z$ direction can be explicitly written as
\begin{equation}
L^2_zW^2_z=\left[\sum_{b_z}\left(n^+_{b_z}-n^-_{b_z}\right)\right]^2,
\label{eq:wn}
\end{equation}
where the summation of $b_z$ runs over all bonds along the $z$-direction.
The number of kinks of worldlines on the $b_z$-th bond 
in the positive (negative) $z$-direction, $n^+_{b_z}\, (n^-_{b_z})$,
is given by
$n^{+}_{b_{z}}=b^\dagger_{i+e_z}b_{i}$
($n^{-}_{b_{z}}=b^\dagger_{i}b_{i+e_z}$)
with site $i+e_z$ being the nearest-neighbor site of site $i$ in
the positive $z$ direction.

In the present system, 
the definition of superfluid density has to be generalized 
to take account of the non-uniform transfer integral~\cite{Rousseau}. 
Allowing for the bond-dependent transfer integral, 
the superfluid density in the $z$ direction is given as 
\begin{equation}
  \rho_s^z =\frac{\langle L^2_z\tilde{W}_z^2\rangle}{2t\beta L_z(L_x^{d-1}-1)},
\label{eq:sfdan}
\end{equation}
where the normalized winding number is  
\begin{equation}
  L_z^2\tilde{W}_z^2=\left[t\sum_{b_z}\frac{\left(n^+_{b_z}-n^-_{b_z}\right)}{t_{b_z}}\right]^2
\end{equation}
with $t_{b_z}=t^*$ along the bonds in the special row of the lattice (denoted by the red line in Fig.~\ref{fig:systems}) 
and $t_{b_z}=t$ along the other bonds,
and $d=2$ (3) in the system of a film (bar) geometry.
Here, we assume that $L_{x}=L_{y}$ in the system of the bar geometry.

\section{Results}\label{sec:results}

\subsection{Film: Anisotropic two-dimensional lattices}\label{sec:film}

\begin{figure}[htpb]
  \includegraphics[angle=0,width=9cm,trim= 0 0 0 0,clip]{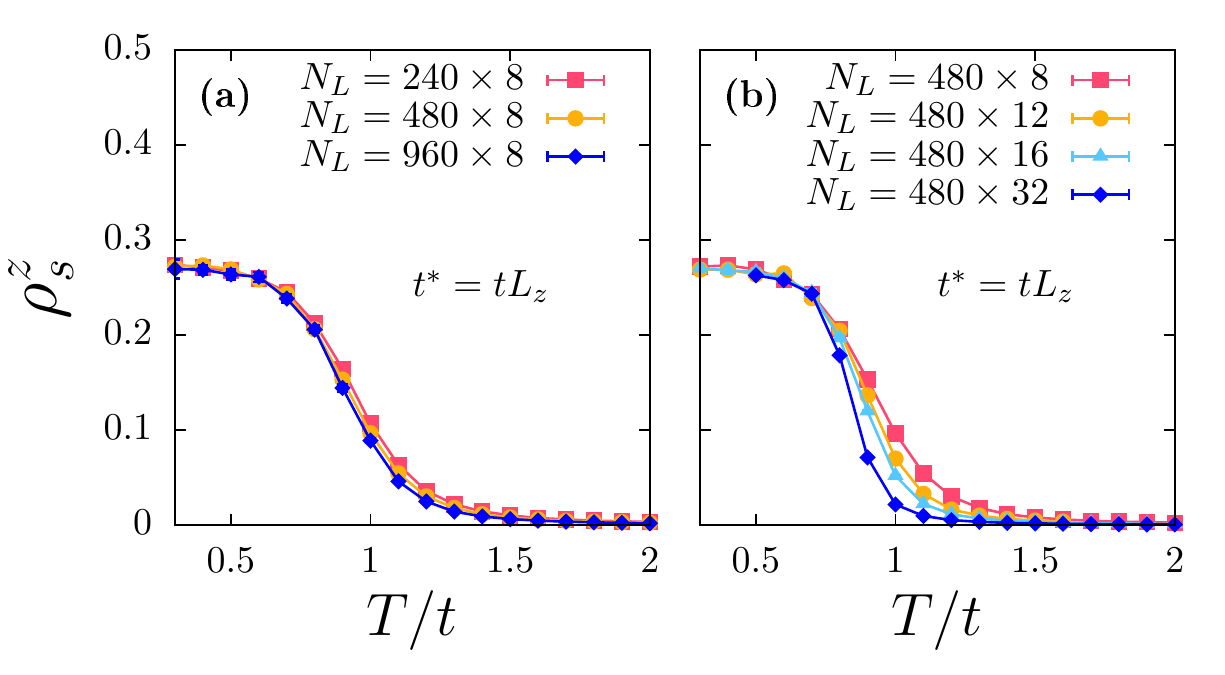}
  \includegraphics[angle=0,width=9cm,trim= 0 0 0 0,clip]{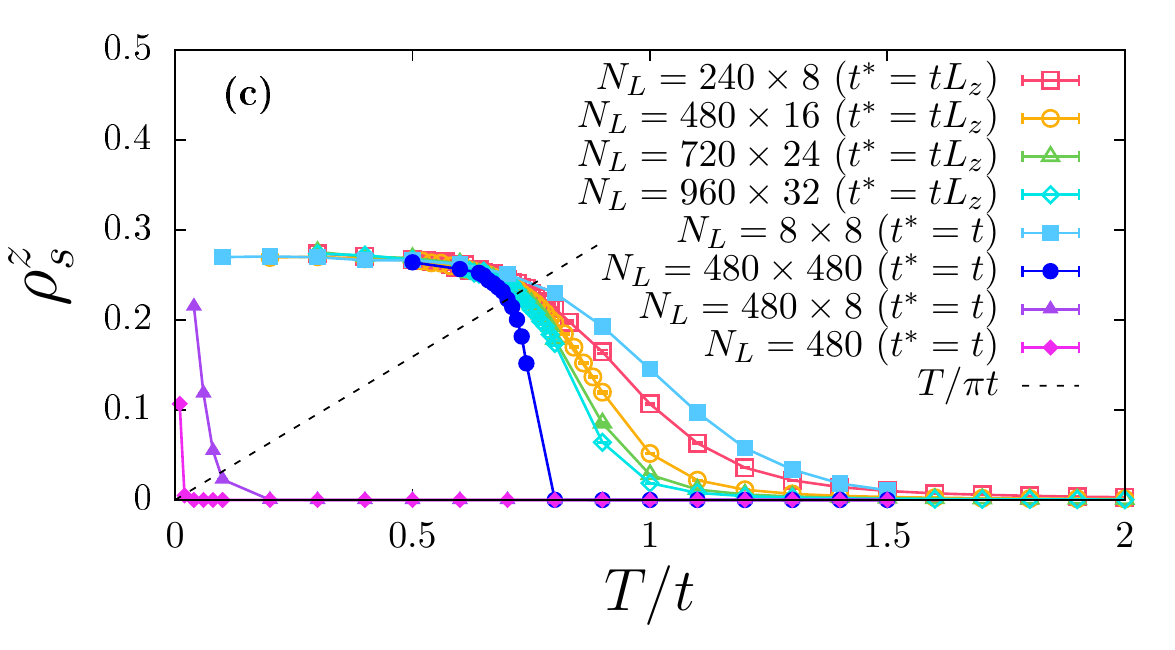}
\caption{
Superfluid density $\rho_s^z$ along the $z$ direction in the film geometry 
of different lattice sizes ($N_{\rm L}=L_z\times L_x$) with $t^*=tL_z$. 
(a) $L_z$ is varied with keeping $L_{x}=8$,  
(b) $L_x$ is varied with keeping $L_{z}=480$, and 
(c) $L_x$ and $L_z$ are varied with keeping the aspect ratio $L_{z}/L_{x}=30$. 
For comparison, 
the results for $N_{\rm L}=8\times 8$, $480\times 480$, $480\times 8$, 
and $480\times 1$ with $t=t^{*}$ are also shown in (c). 
The dashed line in (c) represents $\rho_s^z=T/(\pi t)$. 
The BKT transition temperature $T_{\rm BKT}$ for the two-dimensional system 
determined previously 
by the quantum Monte Carlo method is $T_{\rm BKT}/t=0.68606$~\cite{Melko}. 
}
\label{fig:filmSFDsize}
\end{figure}

Figure~\ref{fig:filmSFDsize} shows temperature
dependence of the superfluid density in the $z$ direction of hard-core bosons
on an anisotropic two-dimensional lattice (i.e., a film) of different sizes. 
When the system is isotropic and large, that is, 
$L_{x}=L_{z}\gg 1$ and $t^{*}=t$,
the superfluid density is found to vanish at $T\simeq 0.7 t$~[see Fig.~\ref{fig:filmSFDsize}(c)], 
which is close to the known results $T_{\rm BKT}/t=0.68606$ in two dimensions~\cite{Melko}.
In the one-dimensional limit i.e., $L_{z}\gg L_{x}$, with $t^{*}=t$, the superfluid 
density vanishes at a much lower temperature than $T_{\rm BKT}$, 
in agreement with the theoretical prediction~\cite{Shev,Machta,SP} and
the previous result for a classical model~\cite{Yama1}~[see the
results for $N_{\rm L}=480\times 8$ and $480\times 1$ in Fig.~\ref{fig:filmSFDsize}(c)].

We then suppress phase slippage by setting $t^{*}=L_{z} t$ to find that
the superfluid density remains finite at finite temperatures. 
In Fig.~\ref{fig:filmSFDsize}(a), $L_{x}$ is fixed at $L_{x}=8$ and
the length $L_z$ of the system is changed. It is observed that the temperature dependence
hardly depends on $L_{z}$ although the aspect ratio $L_{z}/L_{x}$ significantly changes. 
This result clearly shows that the superfluid density remains finite up to $T\simeq t$ 
even in the one-dimensional limit of $L_{z}\rightarrow \infty$ as long as the phase 
slippage is suppressed.

In Fig.~\ref{fig:filmSFDsize}(b), on the other hand, 
$L_{z}$ is kept constant at 
$L_{z}=480$ and $L_x$ is varied. 
As $L_{x}$ increases, the superfluid density is found to drop more
sharply as a function of $T$. As $L_{x}$ approaches to $L_{z}$, the result almost converges to that 
in the two dimensional case for $N_{\rm L}=480\times 480$ with $t^{*}=t$ 
shown in Fig.~\ref{fig:filmSFDsize}(c). 
Figure~\ref{fig:filmSFDsize}(c) shows the superfluid density for a fixed aspect ratio 
$L_{z}/L_{x}=30$. We find that the temperature dependence
of the superfluid density is very similar to that shown in Fig.~\ref{fig:filmSFDsize}(b).

These results clearly demonstrate that the temperature dependence of the superfluid density 
is similar for all cases on the two-dimensional, quasi-one-dimensional, or 
one-dimensional lattice, that is, the superfluid density remains finite at finite
temperatures, provided that the phase slippage is prohibited.
It is also noticed that the temperature dependence
of the superfluid density is primarily determined by $L_x$.

In Fig.~\ref{fig:filmSFDsize}(c), the dashed line represents the universal 
jump of the superfluid density for the BKT transition~\cite{unijump}.
The result for $L_{z}/L_{x}=30$ appears to merge at 
$T\simeq 0.7 t$ to the universal jump line, as the data for $L_x=L_y$ 
(i.e., $N_{\rm L}=8\times 8$ and $480\times 480$) do. 
This strongly suggests that the system undergoes a transition that 
belongs to the BKT universality class. 
In Sec.~\ref{subsec:FSS}, we shall perform the scaling analysis to 
show that the transition is indeed the BKT transition and estimate the 
transition temperature.

\subsection{Bar: Anisotropic three-dimensional lattices}

Now, we study the superfluid density of hard-core bosons on an anisotropic 
cubic lattice composed of $N_{\rm L}=L_{x}\times L_{y}\times L_{z}$ sites 
with $L_{z} \gg L_{x},\, L_{y}$.
As in the case with the film geometry, the superfluid density rapidly diminishes at 
temperatures much smaller than the bulk transition temperature  
$T_{\lambda}\simeq 2t$~\cite{Betts,Peder,Melk,Laflo,Rigol} 
when $L_{z} \gg L_{x}=L_{y}$ and $t^{*}=t$, 
although those results are not presented here.

\begin{figure}[htpb]
\includegraphics[angle=0,width=8.5cm,trim= 0 0 0 0,clip]{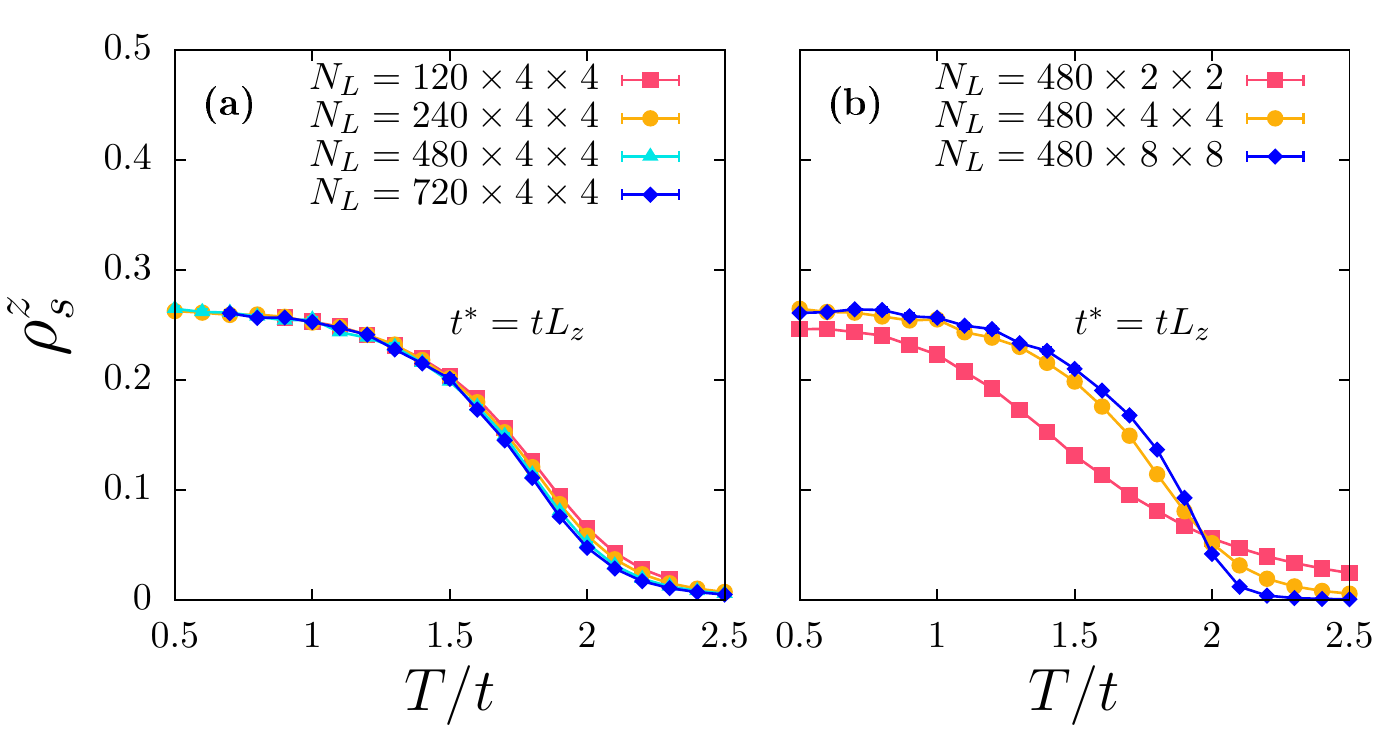}
\caption{
Superfluid density $\rho^z_s$ along the $z$ direction in the bar geometry  
of different lattice sizes ($N_{\rm L}=L_z\times L_x \times L_y$) 
with $t^{*}=L_{z}t$.
(a) $L_z$ is varied with keeping $L_{x}=L_{y}=4$ and 
(b) $L_x=L_y$ is varied with keeping $L_{z}=480$.
}
\label{fig:barSFDsize}
\end{figure}

Figure~\ref{fig:barSFDsize}(a) shows the results of the superfluid density 
for different values of $L_z$ with keeping $L_{x}=L_{y}=4$ and $t^{*}=L_{z} t$ 
to suppress the phase slippage. 
It is observed that the superfluid density is now survived 
up to the bulk transition temperature $T_{\lambda}\simeq 2t$~\cite{Betts,Peder,Melk,Laflo,Rigol}.  
Interestingly, the results hardly depend on the value of $L_{z}$ and remain intact even in the one-dimensional
limit of $L_{z}\rightarrow \infty$.
This is very similar to the results for the film case [see Fig.~\ref{fig:filmSFDsize}(a)]. 
Figure~\ref{fig:barSFDsize}(b) shows the results for different values of $L_{x}=L_y$
with keeping $L_{z}=480$ and $t^{*}=L_{z} t$.  
As $L_{x}$ increases, 
the superfluid density 
vanishes more steeply with $T$. However, it quickly converges in 
increasing $L_x$. 
These results shown in Figs.~\ref{fig:barSFDsize}(a) and 
\ref{fig:barSFDsize}(b)
imply, as in the case of the film geometry, that
the superfluid density remains finite up to the bulk transition temperature
even in the one-dimensional limit of $L_{z}\rightarrow\infty$, as long as 
the phase slippage is prohibited.

\subsection{$t^{*}$-dependence}

Thus far, we have set $t^{*}=L_{z}t$ to suppress the effect of phase 
slippage. However, this value is chosen rather arbitrarily.
Here, we examine the dependence of the superfluid density on the value 
of $t^{*}$ and show that the results do not depend
on the precise value of $t^{*}$ as long as it is large enough 
(i.e., $t^{*} \gtrsim L_{z}t/8$ for $L_{z}=480$).

\begin{figure}[htpb]
\includegraphics[angle=0,width=8.5cm,trim= 0 0 0 0,clip]{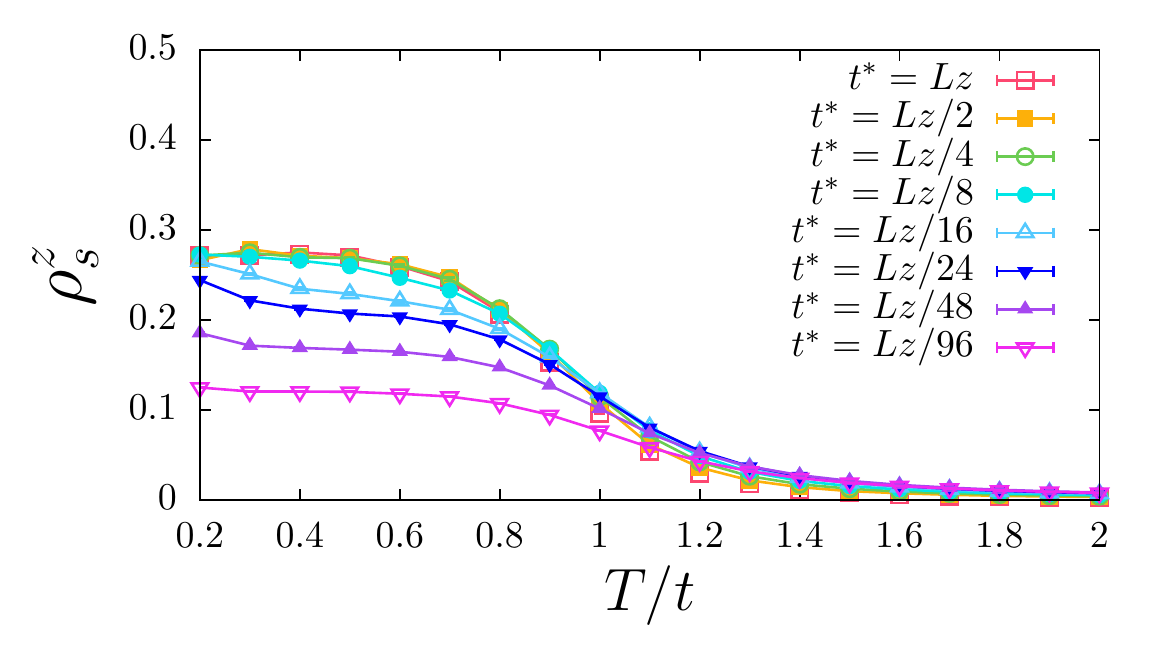}
\caption{
Superfluid density $\rho_s^z$ for different
values of $t^*$ in the film geometry of $N_{\rm L}=480\times 8$ sites with $L_z=480$ and $L_x=8$.
}
\label{fig:filmSFDtx}
\end{figure}

Figure~\ref{fig:filmSFDtx} shows the superfluid density
for different values of $t^*$ in the film geometry of $N_{\rm L}=480\times 8$.
As $t^{*}$ increases, the superfluid density at low temperatures increases, 
because of the suppression of the phase slippage, and 
the results are essentially converged for $t^{*} \gtrsim L_{z}t/8$. 
This implies that the results obtained above are not the results 
for a particular value of $t^*$, but represent the characteristic behavior of
the superfluid density in the systems where phase slippage is suppressed.

\subsection{Finite size scaling for $\rho_s^z$}\label{subsec:FSS}

\subsubsection{Film geometry}

When the system size increases, the system ultimately reaches the 
thermodynamic 
limit irrespective of the shape of the system. 
For the film geometry with a fixed aspect ratio of $R_{\rm film}=L_{z}/L_{x}$, 
we can reach the thermodynamic limit of the two dimensional system as 
$L_{z}\rightarrow\infty$ even when $L_{x}\ll L_{z}$.
However, it is known that $\rho_{s}^{z}$ depends on the aspect ratio $R_{\rm film}$, and 
the temperature where superfluidity sets in decreases from $T_{\rm BKT}$ for $R_{\rm film}=1$ with  
increasing $R_{\rm film}$, when the phase slippage is not 
prohibited~\cite{Melko}.

As shown in the previous section, if the phase slippage is suppressed, 
$\rho_{s}^{z}$ can be clearly finite up to the temperature close 
to $T_{\rm BKT}$ 
for the isotropic two-dimensional lattice with $R_{\rm film}=1$, 
even when $R_{\rm film} \gg 1$  [see Fig.~\ref{fig:filmSFDsize}(c)]. 
Here, using the finite size scaling, we show that the transition 
in the absence of the phase slippage 
is indeed the BKT transition and estimate the transition temperature.

\begin{figure}[htpb]
\includegraphics[angle=0,width=9cm,trim= 0 0 0 0,clip]{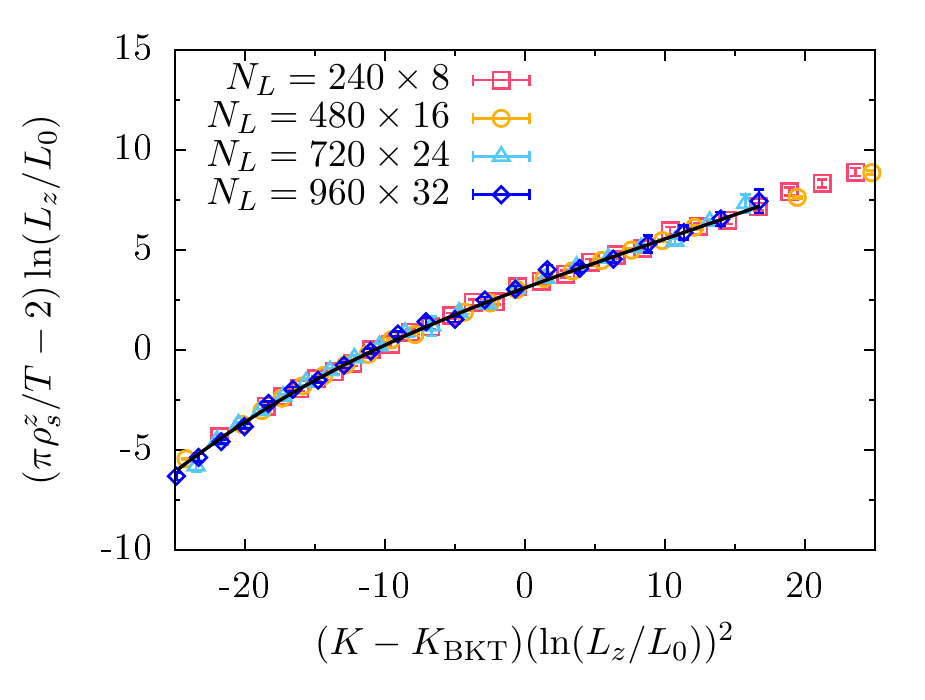}
\caption{
Finite size scaling of superfluid density $\rho_s^z$ in the film geometry with a fixed value 
of the aspect ratio $R_{\rm film}=L_{z}/L_{x}=30$ and $t^{*}=L_{z} t$ for $L_{z}=240, 480, 720$, and 
$960$.
The black solid curve is the scaling function obtained by the
Baysian scaling analysis~\cite{BSA,BSA2}.
}
\label{fig:filmSFDscale}
\end{figure}

Assuming that the transition is the BKT transition, 
we can expect that the quantity 
\begin{equation}
x=\frac{\pi}{2}\frac{2t\rho_{s}^{z}}{T}-2,
\end{equation}
i.e., the deviation of the superfluid density from the universal value at $T=T_{\rm BKT}$, 
satisfies the following finite size scaling equation:
\begin{equation}
x(T,L_{z}) = l^{-1}F\left((K-K_{\rm BKT})l^{2}\right),
\end{equation}
where $K=t/T$ and $l=\ln(L_z/L_0)$ with $L_{0}$ being a
phenomenological constant~\cite{Olsson,Harada}. 
Figure~\ref{fig:filmSFDscale} shows the scaling plot of 
$x(T,L_z)$ for the systems in the film geometry with a fixed value of $R_{\rm film}=30$ 
and $t^{*}=L_{z} t$. We employ the Baysian analysis~\cite{BSA,BSA2} to find
the best scaling function. 
It is clearly observed in Fig.~\ref{fig:filmSFDscale} that the numerical data for different sizes 
collapse excellently onto a universal curve.  
The estimated values are $K_{\rm BKT}=1.49\, (2)$ 
and $\ln L_{0}=-3 \, (1)$.
This confirms that the transition is indeed the BKT transition 
for the systems in the film geometry with no phase slippage allowed. 
The estimated 
transition temperature $T_{\rm BKT}/t=0.671\,(9)$
is to be compared with
the value for the isotropic two-dimensional system,
$T_{\rm BKT}/t=0.68606\, (16)$~\cite{Melko}, where the phase slippage is 
not prohibited.

\subsubsection{Bar geometry}

\begin{figure}[htpb]
\includegraphics[angle=0,width=9cm,trim= 0 0 0 0,clip]{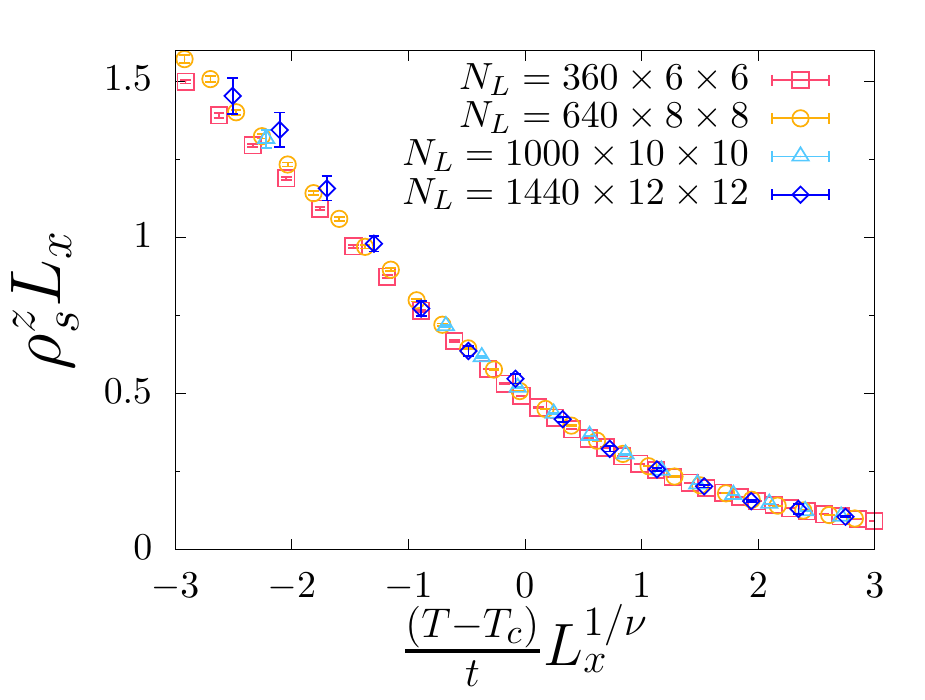}
\caption{
Finite size scaling of superfluid density $\rho_s^z$ in the bar geometry with a fixed value 
of the ratio $R_{\rm bar}=L_{z}/(L_{x}L_{y})=10$ and $t^{*}=L_{z} t$ for $L_{z}=360, 640, 1000$, and 
$1440$. 
Using $\nu=0.6717$, the transition temperature is estimated to be $T_{\rm c}/t=2.022(1)$. 
}
\label{fig:barSFDscale}
\end{figure}

For anisotropic three-dimensional lattices (i.e., bars), the system also 
reaches the bulk limit when we increase $L_{x}$, $L_{y}$, and $L_{z}$
with keeping its relative magnitude constant, $L_{\alpha}=c_{\alpha}L$ 
($\alpha=x,y$, and $z$)
where $L$ and $c_{\alpha}$ are constant, even if $L_{z}\gg L_{x}$
and $L_{y}$. However, the effective length given by
$L_{\rm eff}\simeq L_{z}/(L_{x}L_{y}) = c_{z}/(c_{x}c_{y}L)$
becomes zero as $L\rightarrow \infty$.
Therefore, this limit is rather trivial. 
In contrast, the limit of $L_{z}\rightarrow\infty$ with a constant $R_{\rm bar}= L_{z}/(L_{x}L_{y})$ 
is expected to be nontrivial and
here we discuss the finite-size scaling for the bar systems in this limit.
Figure~\ref{fig:barSFDscale} shows the results for the finite-size scaling of $\rho_s^z$ with $R_{\rm bar}=10$.
We obtained the transition temperature $T_{\rm c}/t=2.022(1)$ in this system
when we use the known critical exponent $\nu=0.6717$ of 3D-XY universality class yielded by classical Monte Caro simulations\cite{Campo, Burovski}.
The transition temperature $T_{\rm c}$ estimated here is shifted from
$T_{\rm c}/t =2.0169\; (5)$ obtained for the half-filled
three-dimensional hard-core boson model~\cite{Rigol} but compared well. 
Therefore, our system in the bar geometry with fixed $R_{\rm bar}$, 
provided that phase slippage is prohibited, is not completely equivalent to the three-dimensional 
isotropic hard-core boson model but belongs to the three-dimensional XY universality class.

\section{Conclusions}\label{sec:conclusion}

We have studied the superfluid density of the quasi-one-dimensional
hard-core boson system focusing on the effect of phase slippage. 
We have demonstrated that the phase slippage is suppressed by setting 
the large transfer integral between the neighboring sites along a single row of the 
system. 
We have successfully shown that superfluid density
remains finite at high temperatures ($T_{\rm BKT}$ in the film geometry 
and $T_{\lambda}$ in the bar geometry) 
even in the one-dimensional limit, $L_{z}\rightarrow
\infty$, as long as the phase slippage is prohibited.
In particular, we have found that the transition in the film geometry is the BKT transition 
if the phase slippage is suppressed.

Although we have shown that the superfluid density can be finite up to 
$T_{\rm BKT}$ or $T_{\lambda}$ in a quasi-one-dimensional system, 
it does not necessarily 
mean that one always observes a finite superfluid density up to those
temperatures in experiments. 
At very low temperatures, i.e., $T \ll T_{\rm BKT}$ or $T_{\lambda}$,
the relaxation time $\tau$ should be long enough compared with the 
inverse of the frequency $\omega$ at which 
the the superfluidity is measured.
Therefore $\omega\tau \gg 1$, and thus $\rho_{\rm s}$ is observed in a torsional oscillator experiment.
As $T$ increases, $\tau$ rapidly decreases and eventually $\omega\tau$
becomes much smaller than unity ($\omega\tau \ll 1$). In this case, it is 
$\rho_{\rm p}$ that is observed with a torsional oscillator. 
However, $\rho_{\rm p}$ readily vanishes at very low temperatures and thus 
$\rho_{\rm p}\simeq 0$ at temperatures where $\omega\tau \ll 1$. 
Thus, in dynamical
experiments, one observes a crossover from finite $\rho_{\rm s}$
to vanishing $\rho_{\rm p}$ (or vice versa)
at a temperature where $\omega \tau \simeq 1$.
The crossover temperature (or the onset temperature) $T_{\rm o}$
increases as $\omega$
increases. What we have found in this study is that the onset temperature 
would be $T_{\rm BKT}$ or $T_{\lambda}$ in the limit of $\omega\tau\rightarrow
\infty$. In other words, the upper limit of the onset temperature is
$T_{\rm BKT}$ or $T_{\lambda}$ in quasi-one dimensional systems.
A crucial point to be emphasized is that the limiting value of the onset temperature 
remains to be $T_{\rm BKT}$ or $T_{\lambda}$  
even in the one-dimensional limit.

If the onset temperature $T_{\rm o}$
is distant from $T_{\rm BKT}$ or $T_{\lambda}$,
we expect to observe a two-step increase in the superfluid density in 
a torsional oscillator experiment. However, if the onset temperature
is close the $T_{\rm BKT}$ or $T_{\lambda}$, it might be difficult to 
separate $T_{\rm o}$ from $T_{\rm BKT}$ or $T_{\lambda}$. In a previous
publication~\cite{Yama2}, 
it was argued that this difference might be the cause for a difference in observation in the film and bar geometries; 
$T_{\rm o} \ll T_{\rm \lambda}$ in the bar geometry, but $T_{\rm o} \simeq
T_{\rm BKT}$ in the film geometry. A more detailed analysis of frequency dependence of 
the superfluid onset is required to clarify this point.

Now, two comments are in order on the direction of future study. First, it
is desirable to extend the present calculation to a continuous system.
It is not clear at all how we can suppress the phase slippage in a continuous
system. A position-dependent mass might be a possible way to suppress the phase slippage. 
Next, it is desirable to calculate the superfluid density observed with
a torsional oscillator in a given system directly under a non-equilibrium condition.  
For this, we further have to specify the microscopic
mechanism of the dissipation of supercurrent, which determines $\tau$, 
such as periodic or random potential caused by the substrate. 
However, currently, sufficient information is not available about such microscopic details of the system.

\section{Acknowledgement}
This work has been supported in part by 
Grant-in-Aid for Scientific Research from MEXT Japan 
(under Grant Nos.~15K05183, 17K14361, and 18H01183). 
The numerical simulations have been performed on 
supercomputers at Supercomputer Center, ISSP, the University of Tokyo, 
and on K computer provided by RIKEN Center for Computational Science (R-CCS) through 
the HPCI System Research project
(Project IDs.:~hp160152, hp170213, hp180098, and hp180129). 

%
%
%

\end{document}